\documentclass{article}
\usepackage{fullpage}
\usepackage{amsfonts,amssymb}
\usepackage{amsthm}
\usepackage{tikz}
\usetikzlibrary{calc}
\usetikzlibrary{arrows}
\usepackage{tikz-3dplot}
\usepackage{color} 
\usepackage[fleqn]{amsmath}
\usepackage{amsthm}
\usepackage{amssymb}
\usepackage{amsfonts}
\usepackage{bbm}
\usepackage{epsfig}
\usepackage{times}
\usepackage{hyperref}
\usepackage{bm}
\usepackage{float} 
\usepackage{appendix} 
\usepackage{lscape}
\usepackage{cite}
\usepackage{mathrsfs}
\usepackage{appendix}
\usepackage{setspace}
\usepackage{mathtools}
\usepackage{authblk}
\usepackage[round]{natbib} 

\theoremstyle{definition}
\newtheorem{definition}{Definition}[section]
\begin{document}

	\title{Determinism Beyond Time Evolution}

 	\author[1]{Emily Adlam} 
	\date{\today} 
	\affil[1]{The University of Western Ontario}

	 \maketitle
	 
	 \abstract{Physicists are increasingly beginning to take seriously the possibility of laws outside the traditional time-evolution paradigm; yet many popular definitions of determinism are still predicated on a  time-evolution picture, making them manifestly unsuited to the diverse range of research programmes in modern physics. In this article, we use a constraint-based framework to set out a generalization of determinism which does not presuppose temporal evolution, distinguishing between strong, weak and delocalised holistic determinism. We discuss some interesting consequences of these generalized notions of determinism, and we show that this approach sheds new light on the long-standing debate surrounding the nature of objective chance.}
	 
	 \newpage

Physicists are increasingly beginning to take seriously the possibility of laws which may be non-local, global, atemporal, retrocausal, or in some other way outside the traditional time-evolution paradigm. Yet many definitions of determinism in usage today are still  predicated on a forwards time-evolution picture, making them manifestly unsuited to the diverse range of research programmes in modern physics. As physics begins to move beyond the time evolution paradigm, is there still a meaningful  notion of determinism as a metaphysical property of the world to be recovered? In this article we argue that there is: we propose a way of thinking about determinism which does not rely on a time-evolution picture, and  we explore some of the consequences of this generalization for the the philosophy of determinism and chance.  

We begin  in section \ref{Determinism} by discussing some existing definitions of determinism. After reviewing some of the flaws of the Laplacean definition, we assess several existing generalisations of the notion. In particular, we note that although the `region-based' formulation used in the study of spacetime theories does avoid building a temporal direction into the definition, this model-theoretic formulation requires us to think of determinism purely as a technical feature of a given theory, rather than a putative  metaphysical feature of the world.  We  therefore consider that there remains a need to understand what sort of metaphysical picture could replace  the temporally directed Laplacean notion of determinism in a post time-evolution paradigm.

In section \ref{det} we use a constraint-based framework to offer several new definitions of determinism in modal terms, distinguishing between strong, weak and delocalised holistic determinism, and we show that these definitions successfully accommodate a range of cases outside the time evolution paradigm. Then in section \ref{app} we discuss some interesting consequences of these generalized notions of determinism. In section \ref{objective}, we show that this approach sheds new light  on the long-standing debate surrounding the nature of objective chance, because it transpires that in a  world satisfying holistic determinism it is possible to have events which appear probabilistic from the local point of view but which nonetheless don't require us to invoke `objective chance' from the external point of view. Finally, in section \ref{rt} we discuss how holistic determinism relates to several other relevant research programmes.

 \section{Definitions of Determinism \label{Determinism}}

 The notion of determinism was most famously articulated by Laplace, who suggested that `\emph{we ought to regard the present state of the universe as the effect of its antecedent state and as the cause of the state that is to follow}.'(\cite{laplace1820theorie}) That is, a theory  is deterministic if according to that theory, the present state of the universe, together with the laws of nature, is sufficient to fix everything that happens in the future. We will refer to this approach as `Laplacean determinism.'

 Although the terminology has evolved somewhat, many modern definitions of determinism are still predicated on the basic idea of the past `determining' the future. For example,   the  model-theoretic definition of determinism, due originally to \cite{Montague1974-MONDT}, tells us that \emph{`a theory is  deterministic if for any two of its models describing the possible temporal evolution of a system, coincidence of the state at one time brings with it coincidence at all times from then on, where “coincidence” is to be spelled out in terms of a suitable mapping.'}. Similarly, \cite{Butterfielddeterminism} suggests that `\emph{[A] theory is deterministic if, and only if: for any two of its models, if they have instantaneous slices that are isomorphic, then the corresponding final segments are also isomorphic.}' And  \cite{doi:10.1093/bjps/axv049} provide a careful classification of three different approaches to defining determinism: we may require that the solution to the differential equations should always be unique, we may require that if two linear temporal realizations  can be mapped at one time they can always be mapped at all future times, or we may require that the theory's models are not branching. As the authors note, all three of their approaches are based on the core idea \emph{`that given the way things are at present, there is only one possible way for the future to turn out'} and therefore all three approaches are still expressions of the temporally directed picture originally associated with Laplacean determinism.

There is an obvious metaphysics underlying this temporally directed picture, which enjoins us to envision the universe as something like a computer, taking some initial state and evolving it forward in time to generate the content of reality(\cite{Wharton}). We can imagine that in Laplace's time this description may have been understood more or less literally. Indeed, as argued by \cite{chen2021governing}, many physicists and philosophers still appear to take this kind of  `dynamic production'  picture seriously - for instance, \cite{Maudlin2002-MAUQNA} advocates a `\emph{metaphysical picture of the past generating the future}.'  And of course, if one has a dynamic production picture in mind then  questions about whether or not  our best current theories obey Laplacean determinism  have profound consequences for our understanding of the kind of world we live in. In particular, proponents of dynamic production will presumably consider that a failure of Laplacean determinism indicates that the universe probably contains some processes which are intrinsically chancy -  for  there must be some  way of deciding what happens when the process of dynamic production arrives at a point where events are not fixed by the laws and the initial state.

 \subsection{The end of dynamic production? \label{td}}

  But there are well-known problems for the Laplacean way of thinking about determinism: in particular,  it fails to properly disentangle the concepts of  determinism and predictability. Laplace's original comments on the matter conflated the two, describing determinism metaphorically in terms of the predictive abilities of \emph{`an intellect which at a certain moment would know all forces that set nature in motion, and all positions of all items of which nature is composed,'}(\cite{laplace1820theorie}) and this way of thinking is still evident in many modern analyses of determinism (\cite{Popper1992-POPTOU-2}). Yet at least  in theory, predictability and determinism are supposed  to be quite distinct - predictability is an epistemic matter, whereas determinism is metaphysical, which is to say, it is supposed to capture something about the way the world really is, independent of our ability to find out things about it. Thus for example \cite{Clark1987-CLADAP-2}, considering the possibility that the functional dependence of the future on the present state might fail to be effectively computable, assert that \emph{`there is no reason at all to tie the claim of determinism to a thesis of global predictability.'} The Laplacean definition doesn't seem to do justice to this intuition.

 Later work has sought to disentangle the notions of determinism and predictability - for example, \cite{Montague1974-MONDT} argues that a purely ontological characterisation of determinism would not necessarily imply deducibility, because deducing the state at a time $t$ from the state at some other time $t_0$ requires us to express those states in sentences, but there are only    denumerably many sentences whereas presumably there are more than denumerably many instants, assuming that instants are in a one to one correspondence with real numbers. And \cite{Earmanbook} writes, \emph{`determinism and prediction need not work in tandem; for the evolution of the system may be such that some future states are not predictable [...] although any future complement than the one fixed from eternity is impossible.'} But as this quote makes clear, much of the relevant work still seeks to understand determinism in terms of forwards temporal evolution (albeit perhaps evolution which we ourselves can't predict), and thus  it does not fully succeed in separating determinism and predictability. For the notion of time evolution is closely linked to prediction: we typically write down laws describing temporal evolution from a present state to future ones, and  non-coincidentally, we ourselves have a strong practical interest in predicting the future from facts about the present. But if deteminism is genuinely to be regarded  as a property of the world rather than as a function of our practical interests, there's no reason it should be defined in terms of (fowards) temporal evolution:  from the point of view of the universe as a whole, it need not be the case that  things are always determined by a kind of process closely modelled on the way in which  in which human observers usually want to predict them. 
 
 Of course, defining determinism in terms of temporal evolution is harmless so long as   it is taken for granted that all laws of nature take  a dynamical,  time-evolution form as suggested by the dynamic production picture. But as argued in refs \cite{adlam2021laws,chen2021governing}, the  dynamic production picture is increasingly under threat in contemporary physics. As a case study, consider   the procedure of Dirac quantisation (\cite{dirac2019principles}) employed in the canonical approach to quantum gravity (\cite{Kucha1993CanonicalQG}). In this procedure, we start with a Hilbert space consisting of all kinematically possible states, and then apply the Hamiltonian and diffeomorphism constraints to arrive at the  physical Hilbert space, consisting of all those states which satisfy the constraints. For example, the Hamiltonian constraint is encoded in the Wheeler-deWitt equation, $H || \psi \rangle = 0$, where $H$ is the universal Hamiltonian and $|| \psi \rangle $ is the universal quantum state. This equation has the consequence that the state $|| \psi \rangle $ does not evolve - that is to say, it can't be regarded as a state at a time, but must be understood as a description of the whole of history at once (this is one manifestation of quantum gravity's famous `problem of time').  Suppose then that we take all of this at face value and we want to consider the constraints as  potential laws of nature. But if we accept that there is no evolution in this picture, clearly they cannot be \emph{evolution} laws. In fact what the constraints do is narrow down the set of physical possibilities to a subset of the space we might originally have envisioned - that is, the constraints  act directly on the space of possibilities. Moreover, clearly if we had enough constraints of this kind, we might eventually find the set of physical possibilities narrowed down to just a single physically possible universe. That, surely, ought to count as a form of determinism: indeed, it would be an example of what  \cite{penrose2016emperor} refers to as `strong determinism,' where the laws of nature single out the course of history uniquely. But the way in which the constraints work is to rule out whole possibilities at once, so there is no guarantee that we will find that given the content of some region of spacetimes we can straightforwardly use the constraints to fill in the rest of the content of spacetime; nor is that a natural thing to try to do in this context, since the constraints do not treat the universe piecewise in this way. These laws, if they are indeed laws, seem to apply to the whole of history `all-at-once,' and so we need a different way of thinking about what it would mean for such laws to be deterministic. 

Indeed, `all-at-once' laws occur even in classical physics. Although it is most common to conceptualise classical mechanics in terms of the Newtonian schema (\cite{Smolinref}) in which laws act on states to produce time evolution, there is also an alternative Lagrangian description of classical mechanics in which systems are required to take the path which optimizes a quantity known as the Lagrangian (\cite{brizard2008introduction}): the Lagrangian approach assigns probabilities to entire histories, rather than taking a state at a time and evolving it forward.  And path integrals - the analogue of the Lagrangian method within quantum mechanics (\cite{feynman2010quantum}) - have become so important to quantum field theory that increasingly we are seeing calls to take the Lagrangian description more seriously (\cite{hartlespacetime, Sorkinpath}). As argued by  \cite{Wharton},  taking Lagrangian methods seriously requires us to adopt an `all-at-once' approach to lawhood in which we think of these laws applying externally and atemporally to the whole of history at once (\cite{Adlamspooky}), so again, we are in need of a different way of thinking about determinism in the context of these sorts of laws. 
 
A multitude of other examples are to be found in contemporary physics. Retrocausal approaches to the interpretation of quantum mechanics have been attracting significant attention in recent years; see for example the two-state vector interpretation (\cite{Aharonov}), the transactional interpretation (\cite{Cramer}), Kent's approach to Lorentzian beables (\cite{Kent}), Wharton's retrocausal path integral approach (\cite{Wharton_2018}), and Sutherland's causally symmetric Bohmian model (\cite{Sutherlandretro}). It is clear that these kinds of models do not fit comfortably into a metaphysical picture of time evolution occurring in only one direction. Likewise,  large sectors of the growing field of quantum information science are concerned with discovering, not evolution laws, but rather general constraints on what information-processing tasks can be achieved using quantum systems. For example, there is the `no-signalling principle'(\cite{MRC}), `information causality'(\cite{Pawlowski}),  and `no-cloning'(\cite{Scaranicloning}).  And the  constructor theory of \cite{Deutsch_2015} expresses physical law not in terms of time evolution, but by appeal to an atemporal characterisation of what tasks are possible and impossible (where \emph{impossible} is understood to mean, not that a process can never occur, but that it can't executed repeatably in a cycle with arbitrary accuracy).  It is not our intention in this article to argue that all of these examples will definitely turn out to be amongst the fundamental laws of the universe, but we do contend that approaches outside the standard time evolution paradigm have become sufficiently prominent in contemporary physics that the philosophy of physics would do well to start addressing them.  And since the Laplacean approach to determinism clearly isn't a useful way of thinking about  constructor laws, retrocausal laws or constraints on information-processing tasks,  there is an obvious  need for a generalisation of determinism which makes sense once we start taking laws outside the time evolution paradigm seriously.  

At this point we should reinforce that it is not our intention to have an argument over terminology - one could certainly make a case that determinism should be defined in terms of evolution forwards in time because that is the way the term has always been used, and proponents of this way of thinking might then be inclined to suggest that worlds governed by laws outside the time evolution paradigm should never qualify as deterministic. We have no particular quarrel with this position: our intention here is simply to argue that there is a meaningful distinction between worlds in which everything is determined by laws which may fail to take a time evolution form,  and worlds in which some events are not determined by anything, so  there is a need for terminology which recognises that distinction. For convenience and familiarity we will  describe the former as `deterministic' and the latter as `indeterministic,' but readers who prefer to use the term determinism to refer only to determination by forwards-evolving states are welcome to adopt whatever alternative terminology they prefer.

 \subsection{Generalisations of Laplacean Determinism} 
 
 One fairly obvious approach to generalising determinism involves allowing evolution backwards as well as forwards - for example, \cite{Maar2019-MAAKOD} distinguishes between systems that are `futuristically deterministic' (their evolution into the future is unique) and `historically deterministic' (their evolution into the past is unique). This is indeed an improvement, in that it no longer insists arbitrarily that determination must occur in one particular temporal direction. But clearly this approach is still predicated on an \emph{evolution} picture, and thus it won't work well for many of the types of laws that we have just discussed. For example, consider a retrocausal world  in which   the initial and final states of the universe together with the laws of nature suffice to fix everything which happens over the whole course of history: neither the initial or final state alone will fix the course of history, so if we write down a theory of this world in a time-evolution form it won't show unique evolution into either the future or the past. And yet there still  seems to be a sense in which this world is not really indeterministic - after all  nothing that happens in it is random and we have no need to invoke anything that looks like an objective chance \footnote{Note that here and throughout this article we will use   `objective chance' to refer exclusively to chances which arise from the fundamental laws of nature, such as the probabilities arising from the Born rule within indeterministic interpretations of quantum mechanics - i.e.  in this article `objective chance' does not include higher-level emergent chances, chances derived via the method of arbitrary functions, deterministic probabilities or anything else that might in another context be called an objective chance.}. This makes it clear that the generalisation we're seeking must involve a more radical conceptual shift than simply allowing for some backwards evolution.

 Another possible generalisation has been proposed by \cite{articleDowe}  to deal with the possibility of worlds in which the relationships between events are temporally non-local. For example, consider a non-Markovian world which is governed by laws such that the evolution at a given time depends not only on the present state, but also on some facts about the past which are not recorded in the present state. The non-Markovian world fails to satisfy  any definition of Laplacean determinism based on states or instantaneous time-slices, since the state at a given time does not suffice to determine all future states. But it seems odd to refer to such a world as `indeterministic' -  after all, the evolution at a given time is entirely determined by the past. Similarly,  consider the example of a world whose ontology contains only pointlike events, so there are no `states' in this world. Clearly in such a world the present state of the universe can't determine the future, since there \emph{is} no present state, but nonetheless it might be the case that the distribution of future events is fully determined by the distribution of past events and under those circumstances one might be inclined to describe this world as `deterministic' even though it doesn't satisfy any definition of Laplacean determinism based on states or instantaneous time-slices,

In response to this problem, \cite{articleDowe} proposes several alternative definitions of determinism -  for example, his modal-nomic definition proposes that a world $W$ is deterministic iff  `\emph{for any time t and any other physically possible world W', if W and W' agree up until t then they agree for all times}.' Both the non-Markovian world and the stateless world can potentially be deterministic according to this definition. But Dowe's approach retains the temporally-directed features associated with dynamic production, and thus it still excludes many types of non-time-evolution laws from consideration.  For example, the retrocausal world where initial and final states fix the course of history will be judged indeterministic by Dowe's criterion. Similarly,  a world governed by `all-at-once' laws in the style of the Wheeler DeWitt equation will   typically be judged as indeterministic by  Dowe's criterion, because in the all-at-once picture, past and future events depend on one another mutually and reciprocally and therefore we will not typically be able to write these dependence relations wholly in terms of determination of the future by the past. Yet `all-at-once' laws could in principle fix the whole course of history uniquely, so it seems very reasonable to think that we ought to have a definition of determinism which allows  retrocausal  and all-at-once worlds  to be deterministic.

 \subsection{Region-based definitions} 
 
 The notion of determinism comes under particular pressure in the context of General Relativity and other spacetime theories, for within such theories it is often challenging to identify what counts as an `instantaneous slice' or even an `initial segment.' Even in special relativity there is no unique notion of a `state at a time,' and when we move to more general cases we have to deal with spacetimes which are not globally hyperbolic and thus have no Cauchy surfaces, i.e. they contain not a single spacetime region which could be regarded as an `initial slice' from which deterministic evolution could take place.   Thus  a generalisation of determinism which seems less closely aligned with the  time-evolution picture has become common in the study of spacetime theories. Roughly speaking, this `region-based' definition says a theory is deterministic if `\emph{agreement on regions of a certain kind (typically sandwiches or slices) forces agreement elsewhere'} (\cite{10.2307/687461}). There has been some disagreement on the precise details of this definition, but here is one example, from \cite{10.2307/687461}: 

\begin{definition} 
 A theory with models $\langle M,O_i \rangle$ is \textbf{S}-deterministic, where \textbf{S} is a kind of  region that occurs in manifolds of the kind occurring in the models, iff: given any two models  $\langle M,O \rangle$ and $\langle M' , O' \rangle $ containing regions S, S' of kind \textbf{S}  respectively, and any diffeomorphism $\alpha$ from S onto S', if $\alpha^{*}(O_i) = O_i'$ on $\alpha (S) = S'$, then there is an isomorphism  $\beta$  from M onto M' that sends S onto S', such that $\beta^{*}(O_i)= O_i'$ throughout M and $\beta(S) = S'$. 
\end{definition} 

Note in particular that this definition of determinism is relativized to a type of region $\textbf{S}$, so we are free to choose whatever kind of region seems most interesting to us - perhaps Cauchy surfaces, if such things occur in the relevant kinds of manifolds, or perhaps some other kind of structure altogether. Since there is no longer an insistence that the determining region should be in the past of the determined region, nor even that the determining region should be of some particular type, this approach does indeed seem to be moving away from the temporally directed dynamic production picture. But what alternative picture does it have to offer? Are we supposed to imagine the universe being provided with the content of the chosen region as an input and then filling in the content of the rest of spacetime? This is not a very compelling picture, not least because if the initial state is not privileged, there is no longer any natural way to identify which particular region we should start from: even in an extremely simple Newtonian spacetime we could start from any time-slice  and generate the rest of history by deterministic evolution forwards and backwards, so if this is to be regarded as a literal description we would seem to have a serious case of underdetermination.  But if we're not supposed to think of the universe as being `produced' from the chosen region, it's unclear what, if anything, we can conclude about reality from the failure of this sort of determinism. Certainly it does not seem to follow that if this sort of determinism fails for some choice of region then the universe must contain processes which are arbitrary or chancy, because one could always make the rejoinder that the content of reality is produced starting from some other kind of region, or indeed reject a `production' metaphysics altogether. 

In fact, the literature seems to suggest that we are as far as possible  to avoid associating any metaphysics at all with this notion of determinism. For example, \cite{Doboszewski2019-DOBRSA-3} contends that `\emph{the issue of determinism cannot be totally decoupled from metaphysical questions ... However one may nevertheless hope to avoid making substantial use of metaphysics in an analysis of determinism.}' And \cite{Muller+2013+47+62} observes that in the model-theoretic approach to determinism,  `\emph{the discussion (of determinism) shifted from the metaphysical question to one about properties of theories.}' The region-based formulation of determinism, being one of these model-theoretic approaches, can only be understood as a property of a given theory; within the model-theoretic paradigm we are actively discouraged from thinking about determinism as a potential property of the world. But if this notion of determinism is not to be regarded as having metaphysical content, what exactly is its purpose? Clearly it is not to be regarded as a practical characterisation of the predictions we can actually make with the given theory: for as we have seen, determinism does not entail predictability, and moreover even in cases where there \emph{would} be predictability in principle,  the definition offers no reason to think that the region $\textbf{S}$ will be one from which we could reasonably collect all the relevant data. Indeed,  \cite{Doboszewski2019-DOBRSA-3} observes that if the region $\textbf{S}$ is supposed to be a Cauchy surface, we will not  even be able to verify the existence of the right type of region in our actual spacetime. Thus it is somewhat unclear why this particular technical notion of determinism should be regarded as an interesting property of a theory; technical features of theories are typically  interesting insofar as they are presumed to reflect some fact about the reality to which the theory corresponds, and/or some practical feature of how the theory works in application, neither of which obviously holds here. 

One might hope to look to the actual practice of physics to understand the utility of the region-based definition. But the main application of the term `determinism' in the General Relativity community appears to be in the context of the initial-value problem, i.e. the problem of determining the course of history given the data on some time-slice $\Sigma$. If the data on $\Sigma$ satisfies certain constraints, then it can be embedded in a unique spacetime which is a solution of Einstein's equations; moreover the spacetime will be globally hyperbolic and  $\Sigma$  will be a Cauchy surface of it (\cite{choquet1969global}). This is important to physicists because under these conditions it's possible to come up with a `time-evolution' version of General Relativity, and also to formulate  canonical quantum gravity (\cite{Kucha1993CanonicalQG}), which is intended to furnish a description of the time evolution of the quantum state of the $3$-metric  (although as we noted earlier, due to the famous `problem of time' it's not entirely clear that the evolution picture succeeds). That is to say, the notion of `determinism' employed by physicists working on General Relativity remains quite closely tied to Laplacean determinism and the time evolution paradigm - and thus, perhaps, involves an unarticulated commitment to a metaphysics of dynamic production. Therefore it's unclear that the justification for this region-based generalisation of determinism can come from within mainstream physics practice.

Moreover,  even if there \emph{are} useful applications for a definition of determinism which is shorn of metaphysical content, we contend that there would also be many useful applications for a definition of determinism which \emph{does} aim to characterise a metaphysically interesting feature of reality. For as \cite{Muller+2013+47+62} puts, it, `\emph{The question of whether the world is deterministic or not, which is perceived by many to have enormous consequences for our understanding of ourselves as free agents, is also first and foremost a metaphysical one ... determinism and indeterminism are metaphysical notions.}' Determinism is frequently invoked in explicitly metaphysical  discussions of topics like  explanation, free will and probability (we will see some examples in section \ref{app}) and it seems unlikely that a definition of determinism which has been deliberately uncoupled from metaphysical commitments will be particularly useful in these applications. Furthermore,  moving away from the time evolution paradigm has interesting consequences for a range of important debates within both physics and philosophy, and in order to fully appreciate these consequences it will be necessary to have some idea of what we might be moving \emph{towards} - that is, we must ask what alternative sort of metaphysical picture might replace the time evolution picture. Understanding the role of determinism in a post-time-evolution setting is surely an important part of that project, and for that we will need something more than the region-based formulation.   
 
In addition to these concerns, the region-based definition of determinism, even considered as a purely technical criterion, becomes difficult to apply in many physically relevant contexts. For example, \cite{Doboszewski2019-DOBRSA-3} points out that when applying this definition to General Relativistic models, the verdict we reach  depends  sensitively on our choice of the type of region $\textbf{S}$: Doboszewski identifies  ten  types of cases which intuitively seem like they should perhaps be regarded as violations of determinism, and observes that there is no particular causal structure that they have in common, so there is no obvious way of identifying a specific type of region which will get the  intuitively `right' verdict in all ten cases. Doboszewski concludes that we should simply adopt a pluralistic approach to determinism, but alternatively one might regard the fact that the region-based definition is so sensitive to a choice of region as evidence that it is perhaps not getting at any particularly interesting or relevant feature of these theories. Even more serious problems are likely to arise in the context of quantum gravity:  for many approaches to quantum gravity treat spacetime itself as emergent from some underlying reality(\cite{cc,Blau2009}), and therefore we will not even be able to identify `regions' in the first place, so the definition of determinism will be applicable at best at an emergent level. This seems troubling - do we really believe that it is conceptually impossible for a theory of quantum gravity to be deterministic in a fundamental sense? 

It's also clear that the region-based formulation will not work well in the context of many approaches outside the time evolution paradigm. For example, consider the retrocausal model where the content of reality is completely fixed by an initial and final boundary condition together, but not by either of them individually. The region-based definition of determinism will allow us to regard this theory as deterministic only if we are willing to accept a `region'  made up of two disconnected parts, which seems to be stretching the definition of the term `region' quite far: and in any case, even if we are willing to accept the union of two disconnected regions as a region, one might well feel that a useful definition of determinism should not make questions about whether a possible world is deterministic depend on the essentially linguistic question of what we're willing to count as a `region.' One possible response to these troubles, as advocated by \cite{Earman2007-EARAOD}, is to adopt a form of quietism, i.e. say that the notion of determinism may simply be inapplicable in many cases that interest us in modern physics. But the alternative is to say that the `region-based' approach is perhaps not the  right generalisation of the original idea of Laplacean determinism, or at least that  the generalisation we propose in this article may be more appropriate in some  cases  - particularly cases where we are concerned not only with technical criteria but with metaphysical content.

\section{The Constraint Framework \label{LON}} 

 Because we are concerned here to analyse determinism as a metaphysical property of worlds, rather than a formal property of theories, we will follow \cite{Muller+2013+47+62} in eschewing    model-theoretic language  and instead characterising the notion  in modal terms, since determinism and objective chance are patently modal notions: \emph{`They are not about what does or doesn’t happen, but about what can or what has to happen and are thus built upon the modal notions of possibility and necessity.'}( \cite{Muller+2013+47+62} ) Our approach to determinism will therefore be   predicated on a fairly robust approach to modality - we will assume that for every possible world there exists  some well-defined modal structure which we can invoke to determine whether or not that world is deterministic. We will not comment further on the nature of that modal structure - it might be defined by the axioms of the  systematization of that world's Humean mosaic which is robustly better than all other systematizations (\cite{lewishumean}), or it could result from governing laws  within any one of the realist approaches to lawhood (\cite{Armstrong1983-ARMWIA,Bird2005-BIRTDC, Ellis2001-ELLSE-2}), or it could be understood in terms of Lewis' modal realism (\cite{Lewis1986-LEWOTP-3}), or it could be something else altogether.

Objective modal structure is most commonly discussed in the context of ontic structural realism (\cite{Ladyman2007-LADETM}). In some cases this modal structure is characterised in terms of \emph{causal} structure (\cite{Berenstain2012-BEROSR,doi:10.1080/02698590903006917}), but as argued in \cite{adlam2021laws}, due to the asymmetrical nature of the standard notion of causation this approach doesn't work well when we're dealing with possibilities outside the time evolution paradigm, so in this paper we will instead adopt the more general approach to modal structure advocated by  \cite{Ladyman2007-LADETM}. In order to formalise this notion of modal structure, we will employ the framework developed in  \cite{adlam2021laws},  which characterises laws of nature in modal terms as \emph{constraints}. This approach is inspired by the increasing prominence of constraint-based laws in physics, as briefly surveyed in section \ref{td}; a very similar account of lawhood in terms of constraints was recently given by  \cite{chen2021governing}.

To formalise the notion of a constraint we make use of the notion of a Humean mosaic, which is understood to consist of all and only the matters of local, non-modal particular fact in a given world.  Obviously what counts as a matter of non-modal particular fact will vary according to one's other metaphysical commitments, and indeed this is our purpose in employing the general term `Humean mosaic' - our aim is to offer a definition of determinism which is compatible with a variety of  accounts of the non-modal content of reality. We emphasize that the use of this terminology is not intended to reflect a commitment to the standard Humean ontology consisting only of the Humean mosaic -  we use the phrase `Humean mosaic' to refer to all of the actual, non-modal content of reality, but we do not intend by that usage to rule out the possibility that reality also has genuine modal content: for example, a realist about modality might wish to maintain that constraints are ontologically prior to the actual Humean mosaic and thus play a role in determining its content.   

Because we don't wish to make our approach to determinism contingent on any particular account of modality,  we will henceforth define constraints \emph{extensionally}, appealing to techniques employed in modal logic: a constraint will simply be defined as a set of Humean mosaics, i.e. the set of all mosaics in which that constraint is satisfied. In some cases, a constraint will be expressible in simple English as a requirement like `no process can send information faster than light,' such that the  constraint corresponds to exactly the set of mosaics in which the requirement is satisfied. But there are many sets of mosaics which will not have any straightforward English characterisation, so we will not be able to state the corresponding constraint in simple terms. Nonetheless, each set still defines a unique constraint.

Using these definitions, we postulate that every world has some set of laws of nature which are an objective fact about the modal structure of that world, and we characterise these laws in terms of \emph{probability distributions over constraints}. Within this picture, we can imagine that the laws of nature operate as follows: first, for each law a constraint is drawn according to the associated probability distribution, and then the constraints \emph{govern} by singling out a set of mosaics from which the Humean mosaic of the actual world must be drawn - i.e. the actual mosaic must be in the intersection of all the chosen constraints. That is to say, the laws of nature associated with a given world are understood to operate by narrowing down the set of physical possibilities for that world, thus dictating what properties the Humean mosaic for that world is required to have. We note  that  we could have given a similar definition in terms of probability distributions over mosaics rather than constraints, but we have chosen to use constraints here in order to acknowledge that a law may require (deterministically or probabilistically) that the actual Humean mosaic belongs to a given set of mosaics, but then say nothing further about which particular mosaic within the set will be selected - that is, the law prescribes no distribution within the set, not even the uniform distribution. This point will be a crucial feature of our approach to determinism.

As shown in  \cite{adlam2021laws}, a very large class of possible laws can be written in this framework. The Wheeler deWitt equation can be analysed as a law assigning probability $1$ to the constraint consisting of all Humean mosaics instantiating a universal quantum state $|| \psi \rangle$ satisfying $H || \psi \rangle = 0$. A  prohibition on superluminal signalling can be analysed as a law assigning probability $1$ to the constraint consisting of all the Humean mosaics in which no superluminal signalling occurs, and $0$ to all other constraints. A constructor law can be analysed as a law assigning probability $\epsilon$ to the set of all mosaics in which some `impossible' process occurs in a repeated cycle, where $\epsilon$ is very close to zero. And given a probabilistic  evolution equation, we can use the equation to construct a probability distribution  $p_s(h)$ over each possible history $h$ of a system $s$, and then analyse the equation as a law which assigns probability $p_s(h)$ to the constraint consisting of the set of all mosaics in which system $s$ has history $h$.  Thus we can use these probability distributions over constraints to characterise the laws of nature without needing to be too specific about what sorts of laws we mean, and this makes it possible for us to give a general definition of determinism which is applicable across many different conceptions of the content of physical law. 

We acknowledge that the idea of a constraint being selected according to some probability distribution may give rise to some scepticism, given that this process presumably only happens once and therefore no relative frequency interpretation is possible. Indeed, we share that scepticism, and would be inclined to adduce the difficulty of making sense of these sorts of probability distributions as a reason to believe that the world  is likely to be deterministic in at least one of the senses described below. But we would argue that any conception of the content of physical reality which holds that there exist genuine objective chances is in effect committed to a probability distribution over Humean mosaics, and therefore if we are to articulate the notion of determinism it is necessary  to make reference to the possibility of probability distributions of this kind in order to characterise the relevant contrast class - whether or not that possibility is actually a coherent one is a further question, and one which we will not have space to address here.

\section{Holistic Determinism \label{det}}

Now our task is to find a way of decoupling the notion of determinism from time evolution. One obvious way to achieve this would be to appeal to the intuitive idea that determinism and objective chance are mutually exclusive: that is, either it is the case that the fundamental laws of nature are deterministic, or it is the case that they involve some sort of `intrinsic randomness.' This line of thought leads to the idea that we should simply define determinism as the absence of objective chance.  But the  problem with this route is that we don't really have a strong grasp on what objective chances are either - the only generally agreed-upon characterisation is that they should obey the Principal Principle (\cite{Lewis1980-LEWASG}). 

The constraint framework provides a precise way of formulating this idea. Indeed, it allows us to disambiguate several different ways in which a world may be deterministic. First, we will say a world (which, recall, can be  associated with some  set of laws of nature) obeys \emph{holistic determinism} iff the probability distributions induced by the laws are trivial: 

\begin{definition} 
	
A world  satisfies \emph{holistic determinism} iff every one of its fundamental laws induces a probability distribution which assigns probability $1$ to a single constraint and zero to all disjoint constraints.
	
	\end{definition}

When this condition is satisfied, we can get rid of the probability distributions altogether and simply assign to each law of nature a constraint, so the actual mosaic must lie in the intersection of all of these constraints. This is the analogue in the constraint framework of the idea that determinism is associated with the absence of objective chance: `no objective chance' is translated as `there are no objective probability distributions over constraints.' Of course the actual mosaic must still be selected from the intersection, but this selection does not need to involve any objective chances, since as noted  in section \ref{LON} the laws do not  define any particular probability distribution over the mosaics in the intersection, not even the uniform distribution.

However, it might be argued that defining determinism as the absence of objective chance is in some cases  too broad. Consider for example a version of classical electromagnetism (\cite{Rousseaux:2005hn}) which includes the stipulation that the electromagnetic potentials are real fields which are a part of nature (and thus they feature in the Humean mosaic) - that is, different configurations of the potentials represent physically inequivalent situations.  Intuitively, it seems clear that this world is not deterministic, since a large variety of physically inequivalent configurations of the electromagnetic potentials will be compatible with the laws in any given situation. But the theory defines no probability distribution over the different configurations of the electromagnetic potentials, so as far as we know  there aren't really any  `chances' here, not even the uniform distribution - some configuration must occur, and so some configuration does, but there is nothing that can be said about how likely different configurations are to occur. We will henceforth refer to cases of this kind, involving events which are not determined by anything but which also are not chancy events, as `arbitrary.'  Thus it seems that a generalized definition of determinism should allow that worlds containing arbitrarinesss of this kind may be indeterministic even in the absence of objective chance.

In light of this possibility, we distinguish further between strong and weak holistic determinism:

\begin{definition} 
	
A world satisfies  \emph{strong holistic determinism} iff it satisfies holistic determinism and there is only one mosaic in the intersection of the set of constraints associated with its laws of nature
	
\end{definition}
 
\begin{definition} 
	
A world satisfies \emph{weak holistic determinism} iff it satisfies holistic determinism and there is more than one mosaic in the intersection of the set of constraints associated with the laws of nature
	
\end{definition}

In the case of strong holistic determinism the actual mosaic is singled out uniquely by the laws and we have neither chance nor arbitrariness; whereas weak holistic determinism allows arbitrariness, but not chance. Thus classical electromagnetism with ontological electromagnetic potentials will satisfy weak holistic determinism but not strong holistic determinism.  One might perhaps question whether `weak holistic determinism,' really ought to count as a form of determinism, given that it allows events to occur which are not determined by anything. However, even worlds satisfying Laplacean determinism exhibit some arbitrariness - typically in such a world the initial state of the universe is `arbitrary' in just the same way as the configuration of the electromagnetic potential, since we don't assign objective chance distributions over initial states but we don't usually take them to be  determined by anything either. So when we set out to generalize determinism in a way that does not give any special status to the initial state of the universe, we have two options - either we allow arbitrariness elsewhere as well, or we allow no arbitrariness whatsoever. Weak holistic determinism takes the former route and strong holistic determinism the latter, so both have at least some claim to be the spiritual heir of Laplacean determinism.  

That said, there does seem to be a sense in which a world satisfying Laplacean determinism is more strongly deterministic than electromagnetism with ontological electromagnetic potentials,  since the arbitrariness involved in the former is limited to the start of time, whereas in the latter, even if we know the full content of spacetime except for some small region, we still won't be able to determine the electromagnetic potentials in that region (since it is always possible to perform a gauge transformation on the potentials which goes to zero outside the region).  So ideally we would like to offer some further definition which distinguishes between worlds satisfying Laplacean determinism and worlds like the one governed by electromagnetism with ontological electromagnetic potentials. Of course, we could add a special exception for the initial state into our definition of determinism, but this would seem like a relic of the older temporally directed approach - if we're no longer insisting that laws must take the forward time-evolution form there's no good reason to treat the initial state as special. So instead we suggest the following definition:

\begin{definition} 
	
	A world associated with a set of laws of nature satisfies \emph{delocalised  holistic determinism} iff it satisfies holistic determinism,   and there is no pair of mosaics in the intersection of the set of constraints associated with its laws of nature which are identical everywhere except on some small subregion of spacetime

\end{definition}

 Note that this definition assumes the existence of a natural map between the two spacetimes associated with two different mosaics.  Of course in the case where two mosaics are associated with spacetimes having very different structure it may not be the case that there exists a unique natural map between their spacetimes (although perhaps such a thing could be arrived at using the best-matching techniques of \cite{articleBar}). However, in that case clearly we would not want to say that the two mosaics are identical everywhere except on some small subregion of spacetime, so there is no need to invoke a map; this definition becomes relevant only when we are dealing mosaics with spacetimes similar enough to support a natural map between the spacetimes, except possibly on some small subregion.
 
 The point of this  definition is that it  ensures the arbitrariness involved in selecting the actual mosaic can't be localised in any particular region, which means we can't have any indeterministic `holes.' To see how the definition works, consider a world satisfying Laplacean determinism in which the laws of nature are time-reversal invariant. In the constraint framework, this world has a set of laws of nature which are each associated with a single constraint, and the intersection of these constraints is a set of mosaics such that each mosaic in the set has a different initial condition. Moreover, because the laws of nature of this world are deterministic and time-reversal invariant, it must be the case that no   pair of mosaics in the set are the same on any time-slice after the beginning of time, since otherwise reversing the direction of time would take a single state into two different states, in violation of the assumption that the laws of this world satisfy Laplacean determinism. Thus when the actual mosaic is drawn from this set of mosaics, this has the effect of fixing the state of the world on the initial state, but since the mosaics differ at all subsequent times as well, one could equally well regard this process as fixing the state of the world on any other time slice. So there is no fact of the matter about which \emph{particular} time-slice is responsible for the rest of history: when we select one of the allowed mosaics we determine all the time-slices at once. Therefore the arbitrariness involved in a  universe satisfying Laplacean determinism can't be located in any specific spacetime region, so the Laplacean universe does indeed satisfy weak delocalised holistic determinism.

 On the other hand, a world governed by electromagnetism with ontological electromagnetic potentials will not satisfy delocalised holistic determinism, because given some mosaic in the intersection of the constraints, we can generate another mosaic which is identical except on some small region of spacetime by simply performing a gauge transformation which goes to zero everywhere outside the region in question, and since the laws don't assign any probabilities over gauge-related configurations, this other mosaic must also lie in the intersection of the constraints.  Thus this definition captures what is distinctive about worlds satisfying Laplacean determinism without needing to give special significance to the initial condition - the important thing about such worlds is not that the arbitrariness occurs only at the start of time, but that the arbitrariness can never be localised anywhere, and thus in general such worlds have a high degree of predictability and consistency even though the laws of nature do not fix everything in the world uniquely.

Evidently all of these definitions are suitable for studying worlds with laws outside time evolution paradigm: worlds with non-Markovian laws, worlds with an ontology that does not include states, worlds with all-at-once laws and worlds with retrocausal laws are all capable of satisfying weak, delocalised and/or strong holistic determinism,  depending on the specific details of the laws in question. The definitions encapsulate   quite a different way of thinking about what it is for a world to be deterministic: rather than treating the initial conditions (or the conditions of some spacetime region) as an input and asking if the input is sufficient to determine the rest of the course of history, we have the laws induce constraints which select entire mosaics at once,  and then we ask if these laws involve any objective chances and if they suffice to pick out the course of history uniquely. Thus these definitions also realise our desideratum of making a clean distinction between determinism and predictability, as they characterise determinism entirely in terms of objective modal structure, without explicitly or implicitly leaning on facts about the practical interests and epistemic limitations of human observers. A world which is deterministic in the holistic sense may indeed exhibit a high degree of predictability, but it also may not - the way in which the laws determine the course of events may be highly non-local in space and time, meaning that it may be impossible for limited local agents to gather enough data to see the whole picture. Thus this framework fulfils the mandate of providing a precise way of thinking about determinism as a metaphysical property of a world rather than a formal property of a theory, thus  opening the door for that notion to be applied to a variety of ongoing metaphysical discussions.

\subsection{Other Comments} 
\begin{enumerate} 

\item The distinction we have made here is similar to one employed in \cite{penrose2016emperor}, which distinguishes between determinism, i.e. what we have referred to as Laplacean determinism (\emph{`if the state of the system is known at any one time, then it is completely fixed at all later (or indeed earlier) times by the equations of the theory'}) and strong determinism, i.e. what we have referred to as strong holistic determinism (\emph{` it is not just a matter of the future being determined by the past; the entire history of the universe is fixed, according to some precise mathematical scheme, for all time.'}). However, Penrose's scheme does not seem to accommodate the intermediate possibility of worlds which satisfy weak holistic determinism but not Laplacean determinism, such as worlds which have some degree of arbitrariness at some point other than the initial state of the universe but nonetheless have no objectively chancy events. Moreover, Penrose is rather vague about what it would take for the entire history of the universe to be fixed, but the constraint framework allows us to be precise: the history of the universe is fixed if and only the laws of nature induce trivial probability distributions over constraints, and the intersection of the constraints associated with these laws contains exactly one mosaic. 

\item In one sense, weak holistic determinism, strong holistic determinism and delocalised holistic determinism are all weaker than Laplacean determinism, since worlds with laws outside the time evolution paradigm like those discussed in section \ref{td} could satisfy any of these forms of holistic determinism without satisfying Laplacean determinism. However, in another sense strong holistic determinism is much stronger than Laplacean determinism, because Laplacean determinism allows for the existence of arbitrariness (in the selection of initial conditions and the selection of parameter values) while strong holistic determinism insists that everything is fixed once and for all by the laws of nature.

\item We reinforce that all of these definitions for determinism depends crucially on the claim that reality has a unique, observer-independent modal structure.  Given any Humean mosaic we could come up with an infinite number of  sets of constraints whose intersection contains only this mosaic - trivially, for a mosaic $A$ we can pick two random mosaics $B$ and $C$ and choose the constraints $\{ A, B\}$ and $\{ A, C\}$. So simply finding such a set which is consistent with our observations doesn't tell us that the world must be deterministic: we must make the further claim that these constraints are \emph{in fact} the constraints which feature in the objective modal structure of reality, and therefore we must believe that there is some fact of the matter about what this objective modal structure really is. So for example, assessing a claim that the world satisfies strong holistic determinism has several steps: first we must verify that the proposed constraints do indeed single out a unique Humean mosaic,  second, we must verify that the proposed constraints are consistent with our observations, and third, we must decide whether the proposed constraints seem like plausible candidates for laws. This last part of the process will naturally be shaped by whatever expectations we have about laws - for example, if we think the laws of nature are simple then we will tend to expect that the constraints  induced by the laws can be expressed in a closed form in simple terms.

\item  Our definitions also depend crucially on the stipulation that the laws of nature assign no probability distribution over the mosaics in the constraints that they induce, nor over the mosaics in the intersection of all of those constraints. However, this certainly does not entail that we ourselves may not assign any distribution over mosaics in the intersection. For we may   have Bayesian priors which lead us to assign some distribution over mosaics before taking any of the laws of nature into account - for example we might simply assign the uniform distribution, or  we might choose a distribution which favours simpler mosaics. Thus when we learn about a constraint induced by a law, we will update our beliefs by eliminating all mosaics not consistent with the corresponding constraint and renormalizing our credences, so after conditioning on all of the laws we will be left with a normalized probability distribution which is non-zero only on mosaics in the intersection of the constraints.  In the case of strong holistic determinism, no matter what priors we start out with, after conditioning on the laws we will be left with a distribution that assigns probability $1$ to a single mosaic; but in the case of weak holistic determinism we will in general end up with a non-trivial distribution, which will be the uniform distribution if the original prior distribution was the uniform distribution, but which may be non-uniform if our original priors were non-uniform.
But this distribution is a \emph{subjective} probability distribution derived from our original subjective priors, not an objective chance distribution arising from the laws themselves: so in this case we still have determinism in the sense of `the absence of objective chance,' even though observers may still assign non-trivial probabilities over mosaics compatible with the laws.  This is precisely analogous to the way in which  we might assign subjective probability distributions over initial conditions in the context of Laplacean determinism, for example in analyses of statistical mechanical systems based on the principle of indifference (\cite{Meacham2010-MEACAT}).

\end{enumerate}

   \subsection{Humean Mosaics}

None of our definitions are intended as a direct replacement of  model-theoretic definitions of determinism, because our approach does not simply take a theory  and decide algorithmically whether that theory is deterministic. For a start, the definition applies to worlds, not theories, so to apply it to a theory  (expressed either as a set of models or a set of sentences) we would have to postulate a world governed by that theory, which would require us to translate the theory into putative `laws of nature' in the constraint formulation, and  typically there will be more than one way of doing that. We would also have to decide whether we need to add additional laws of nature from outside the theory, since our definitions require us to consider the complete set of laws of nature associated with a given world.  So in circumstances where one wants nothing more than a simple judgement as to whether some particular \emph{theory} is   deterministic, the   model-theoretic approach is obviously more suitable.  This is not a weakness of our approach, because it was never our intention to define determinism as a formal property of theories. It would be interesting in future work to develop formal expressions of this approach to determinism which can be applied within a model-theoretic framework, but that is beyond the scope of the present project. 
    
Applying our definitions to a specific theory  would also require  a judgement about what sorts of categorical  properties feature in the set of Humean mosaics associated with the theory.  For example, consider quantum mechanics: if we suppose that quantum mechanics is a complete theory of reality and we take it that wavefunction collapse is a part of the Humean mosaic, then the theory does not satisfy weak or strong holistic determinism, whereas if we say that only the unitarily evolving quantum state is part of the Humean mosaic, then we get the Everett interpretation (\cite{Wallace}), which satisfies   weak holistic determinism. Note that the Everett interpretation in its usual form does not satisfy strong holistic determinism, since the initial state of the universe is arbitrary; however, the Everett interpretation coupled with something like Chen's `quantum Wentaculus,' which features a lawlike initial state (\cite{chen2018quantum}), could potentially satisfy strong holistic determinism.

One might worry that this ambiguity is a problem for our approach. However, it is in fact  a deliberate choice, because our intention is to define in a general, metaphysical sense what it means for a world to be deterministic, and therefore it is important to avoid predicating the definition on specific ontological assumptions. Moreover, this definitional quirk is not  unique to our approach  - as emphasized by  \cite{EarmanGR}, judgements about whether a theory satisfies Laplacean determinism  likewise depend on our presuppositions about the possibility space that we're dealing with. Thus, for example, Laplacean determinism is  similarly noncommittal about quantum mechanics. Moreover, this ambiguity has not prevented Laplacean determinism from yielding generally meaningful results in many cases, although of course there are a few interesting grey areas and ongoing debates (e.g. see \cite{EarmanGR, delsanto2021indeterminism}). We are happy for our definition to inherit this particular form of ambiguity - it would not make sense for our definition of determinism to be stricter than Laplacean determinism in this regard, since our intention here was to \emph{generalize} Laplacean determinism and therefore it is important for us to be able to recognise worlds satisfying Laplacean determinism as a subclass of worlds that are deterministic in our sense.

  \subsection{Examples: Spacetime theories}
  
In order to show that the proposed approach to determinism captures some of the intuitive content that is not well addressed by the region-based formulation, we discuss how some of the problem cases  cited by \cite{Doboszewski2019-DOBRSA-3} can be accommodated by the constraint approach.

  \paragraph{Spacetimes violating weak causality conditions} 
  
  Spacetimes violating weak causality conditions do not allow for an initial value problem or anything remotely like one. This means we can't easily apply a region-based definition of determinism, or any other approach which asks us to take some physical data and construct the rest of the solution from it. However, since constraints pick out whole solutions at once, we can certainly have constraints which contain Humean mosaics corresponding to spacetimes which violate weak causality conditions, and therefore we can certainly postulate laws of nature which give rise to such spacetimes, and then check if those laws exhibit one or more of the forms of holistic determinisms we have postulated. It is clear that there do exist laws of nature which will produce such spacetimes, because the Einstein equations can produce them in special cases; and since the Einstein equations in their standard form are `all-at-once' laws whose solution is an entire course of history rather than a state at a time, they are exactly the kind of law for which the constraint formulation is particularly well-suited. 
  
  \paragraph{Spacetimes with closed timelike curves}

 \cite{QMG, adlam2021tsirelsons}  studied the compatibility of `closed loops' with a concept of `global determinism' which is similar to the notion of holistic determinism that we have used in this article.  \cite{QMG, adlam2021tsirelsons}  argues that  the values of variables in these closed loops `come out of nowhere,' which is to say, they are not determined by anything, so closed causal loops  are not consistent with a  universe satisfying holistic determinism. Using the definitions set out in section \ref{det}, we can make this observation more precise. First, clearly variables which are not determined by anything are incompatible with a universe that obeys strong holistic determinism, so the argument of \cite{QMG, adlam2021tsirelsons} work perfectly well if `global determinism' is intended to mean `strong holistic determinism.' Moreover, indeterministic closed loops are also ruled out by delocalised weak holistic determinism, because the arbitrariness associated with the variables in a closed causal loop is localised to the events inside the loop. On the other hand indeterministic closed loops are not ruled out by weak holistic determinism, since the values of variables inside the loop could be arbitrary rather than chancy.  
   
  \paragraph{Extendible Spacetimes}

  An extendible spacetime is a spacetime which is a subset of a larger spacetime(\cite{2012caus}).  Suppose that a given world has a set of laws such that the intersection of the constraints associated with the laws contains a mosaic corresponding to an extendible spacetime. Then, assuming that there is no law specifically prohibiting extensions, it necessarily follows that the intersection of the constraints associated with the laws will contain more than one mosaic, since we can take the mosaic in this intersection which has an extendible spacetime and obtain another mosaic which will also be in the intersection  by extending it. Therefore a world with laws allowing extendible mosaics will typically fail to satisfy strong holistic determinism. Moreover, the extended mosaic will differ from the original mosaic only on a small subregion of spacetime, in the sense that the new mosaic postulates a small region of spacetime that does not exist at all in the original mosaic, and therefore a world with laws allowing extendible mosaics will typically fail to satisfy delocalised holistic determinism. However a world with such laws may satisfy weak holistic determinism, since the laws need not define any probability distribution over the extent of spacetime.  
  
  \paragraph{Spacetimes with holes} 
  
A spacetime is hole-free if  `\emph{the Cauchy development of any spacelike surface is “as large as it can be” '}(\cite{pittphilsci4380})- i.e. it is not the case that the domain of dependence of some partial Cauchy hypersurface can be extended by imbedding it into a different spacetime (\cite{2012caus}). Suppose that a given world has a set of laws such that the intersection of the constraints associated with the law contains a mosaic corresponding to a spacetime which is not hole-free. Then, assuming that there is no law specifically prohibiting extending the domain of dependence of partial Cauchy hypersurfaces, it follows that the intersection of  the constraints associated with the laws will contain more than one mosaic, since  since we can take the mosaic in this intersection which has an extendible spacetime and obtain another mosaic which will also be in the intersection  by extending the domain of dependence of some partial Cauchy hypersurface.  So a world with laws allowing spacetimes that are not hole-free will typically fail to satisfy strong holistic determinism.  Moreover, the extended mosaic will differ from the original mosaic only on a small subregion of spacetime, in the sense that the new mosaic postulates a small region of spacetime that does not exist at all in the original mosaic, and therefore such a world will typically not be compatible with delocalised holistic determinism either. However, a world with such laws may satisfy weak holistic determinism, since the laws need not define any probability distribution over the size of the domain of dependence.

\section{Applications \label{app}} 

We have criticized Laplacean determinism for being  inappropriately intertwined with our practical interests (in particular, our desire to predict the future based on facts about the present); but one might argue that being aligned with our practical interests is precisely what made Laplacian determinism a useful concept, and thus one might worry that generalizing determinism as we have done here serves only to make the concept less useful. Thus in this section we will explain how this generalization of determinism may have important consequences for scientific and philosophical thinking. The concept of Laplacean determinism has played a range of different intellectual roles in science and philosophy, so we will proceed by examining how a generalised notion of determinism may contribute to some of these domains of application. 
 
\subsection{Scientific Explanation}

Determinism has played an important role in shaping our expectations for scientific explanation -  for a long time it was standard in science to explain events by showing how they could be deterministically produced by past conditions. This is formalised in the `deductive-nomological' model for scientific explanation (\cite{popper2005logic,doi:10.1086/286983}), which posits that a valid explanation should be composed of a set of sentences from which the explanandum logically follows, where at least one of these sentences must express a law of nature and that law must be an essential element in the deduction of the explanandum.  Most common examples of DN-explanations involve time-evolution laws or at least temporally directed laws, and thus DN explanations usually involve postulating some past conditions and showing that the deterministic forwards time-evolution of these conditions gives rise to the explanandum. This model is sometimes relaxed to allow for explanations involving statistical inference rather than exact logical deduction (\cite{hempel1965aspects}), but deductive-statistical and inductive-statistical explanations are simply a generalisation of  forwards-evolving deterministic explanations and thus they typically preserve the expectation that the future will be explained by lawlike evolution from the past.

Holistic determinism opens up new possibilities for scientific explanation - for if the definition of Laplacean determinism is largely a function of our practical interests then there is no reason to expect that all valid scientific explanations will be predicated on forwards time evolution, since for scientific realists one of the main purpose of scientific explanation is to achieve \emph{understanding}, to which practical concerns are only tangentially relevant. So when we come upon phenomena  for which there seems to be no satisfactory explanation within a forward time-evolution picture, we should be open to the possibility of explanations based on global laws using the constraint framework. A good  example of this is the past hypothesis: by definition nothing evolves into the initial state of the universe, so the initial state can't be given a lawlike explanation if we assume that laws always take the forward time-evolution form.  But the past hypothesis can  straightforwardly be written as a constraint - it is simply the set of all Humean mosaics in which the arrangement of local matters of particular fact at one end of the mosaic has low entropy (or some other more sophisticated characterisation of the desired initial state - see \cite{pittphilsci8894}). The past hypothesis can therefore be explained in this framework by hypothesizing that there is some law of nature which induces this particular constraint - perhaps something like the `Mentaculus' of \cite{doi:10.1142/9789811211720_0001} or the quantum `Wentaculus' of \cite{chen2018quantum}) - thus ensuring that the initial state will necessarily be a low entropy state whilst the selection of one low entropy initial state in particular remains arbitrary.

Similar arguments can be made about parameter values: for example, cosmologists are currently much exercised over the problem of why the cosmological constant is so much smaller than their models suggest it ought to be (\cite{RevModPhys.61.1}). It's hard to see how this feature could be explained within the kinematical/dynamical picture, because the cosmological constant is usually considered to be a fundamental constant of nature which has always had the value that it has, and so there is no option to say that it has been produced by evolution from some earlier conditions. Thus most current approaches to explaining the value involve fairly exotic explanatory strategies, such as anthropic arguments in the context of a putative multiverse (\cite{Banks:2000pj}). However, within the constraint framework there is no reason why we can't simply explain the cosmological constant in a lawlike manner: all we need to do is suggest that there is some law of nature which induces the constraint consisting of all the Humean mosaics in which the cosmological constant is very small. In the constraint framework, this nomological explanation has exactly the same status as more familiar nomological explanations, such as  explaining that apples fall due to the law of gravity: the key point is that nomological explanations do not necessarily have to have a temporal character. 

Of course,  some  caution is required with this strategy. Seeking explanations for phenomena that we find surprising or conspiratorial is frequently a good way of making scientific progress, so we don't want to make explanation too cheap: answering every scientific question with `Because there's a law which makes it so,' isn't likely to lead to any new insights.  But of course, the same is true with standard DN and inductive-statistical explanation - not every proposed DN or inductive-statistical explanation is interesting and informative, so we have a variety of criteria which can be applied to judge which explanations have merit. These criteria, including simplicity, unification, the absence of fine-tuning and so on, can equally well be applied to constraint-based explanations. Thus rather than simply imposing a constraint consisting of Humean mosaics in which the initial state is simple or a constraint consisting of Humean mosaics in which the cosmological constant is small, we might want to come up with some more general feature from which these constraints could be derived - for example, one could imagine deriving the past hypothesis  from a more general constraint that singles out Humean mosaics which are sufficiently interesting or varied. This constraint rules out mosaics where everything is in thermal equilibrium throughout the whole of history, which does indeed entail that the initial state must have low entropy, but it may also have \emph{other} interesting consequences, and if the consequences turn out to be powerful enough we would have good reason to accept the proffered explanation. The overarching point is that holistic determinism offers us a new sort of explanatory framework: there can be good and bad  explanations within that framework, and there's work to be done to establish the appropriate criteria for judging these sorts of explanations, but nonetheless this new approach is a promising route to answering questions which seem intractable within standard explanatory paradigms. 

\subsection{Assessment of Theories} 

Determinism has long functioned as  a gold standard for an ideal scientific theory - for example, the idea that quantum mechanics might be indeterministic was accepted only begrudgingly by many physicists at the time of the theory's formulation, as witnessed by Einstein's complaint that `He (God) does not play dice.'(\cite{doi:10.1063/1.1995729}) Over the last century many attempts have been made to  `complete' quantum mechanics so that it obeys Laplacean determinism after all (\cite{genovese2005research}), and indeed two of the most popular interpretations of quantum mechanics (the de Broglie-Bohm interpretation (\cite{holland1995quantum}) and the Everett interpretation (\cite{Wallace}) do satisfy Laplacean determinism.

However, if we accept that the standard definition of Laplacean determinism is to some degree a function of our practical interests, there is less justification for regarding Laplacean determinism as the ultimate goal of a scientific theory. Obviously, theories which satisfy Laplacean determinism are still desirable due to their practical utility for predicting the future, but from the point of view of the scientific realist with an interest in understanding how things \emph{really} are, theories which satisfy weak, strong or delocalised determinism may be just as well motivated. After all, all three of these notions of determinism do justice to Einstein's intuition that `He does not play dice,' in the sense that none of them allows the existence of genuine objective chances from the external, `god's-eye' point of view. 

Thus this approach to determinism suggests new ways of thinking about what a good scientific theory should look like. For example, instead of coming up with theories which postulate a set of states and a set of differential equations, perhaps we should be looking more seriously at theories postulating laws which govern `all-at-once' from an external point of view. As noted in section \ref{td}, modern physics does already contain examples of such laws, but in general these examples have been arrived at by starting from a temporal evolution picture and subsequently generalizing it or reformulating it; things might look quite different if we started from the assumption that we are looking for a holistic, `all-at-once' theory and then proceeded without insisting on the existence of a time-evolution formulation of the theory.  Adjusting our gold standard to more closely reflect the form of the true laws of nature is likely to be a good way to stimulate progress toward understanding those laws, so we have strong practical motivations for rethinking the gold standard. 

\subsection{Free Will}

Laplace's ideas about determinism gave rise to a spirited debate over the possibility of free will in a deterministic universe which has continued to this day (\cite{Williams1980-WILFWA, Inwagen1975-INWTIO, Mickelson2019-MICTPO-41}). The possibility of holistic determinism certainly raises new questions in this debate. For example, even if you believe that we don't have free will in the context of Laplacean determinism, you might still be willing to say that we could have free will in the context of some sorts of \emph{holistic} determinism. After all, if the course of history is determined by laws which apply to the whole of history all at once, our actions are determined by other events, but also those events are partly determined by our actions, since each part of the history is optimized relative to all the other parts. So there is a degree of reciprocity which is lacking in the Laplacean context, where the present state determines our actions and our actions do not act back on the present state. This is an interesting direction of enquiry, but somewhat outside the scope of the present paper, so we will leave it as a topic for future research. 

\subsection{Initial Conditions} 

The dynamic production picture on which Laplacean determinism is predicated encourages us to take a particular attitude to initial conditions, regarding them as freely chosen inputs to an otherwise fully determined system. Thus in particular the Laplacean picture makes a sharp distinction between the freely chosen initial conditions and the values of any parameters which enter into the theory (e.g. the masses of fundamental particles, the gravitational constant and so on), which are normally taken to to be nomic (\cite{pittphilsci15757}).  But in the constraint framework the case for this different status is significantly weaker. If the universe satisfies strong holistic determnism then both the initial conditions and the numerical values of the parameters must be determined by the laws of nature - that is, we must have state rigidity and parameter rigidity, in the terms of \cite{pittphilsci15757}. But if the universe is only weakly deterministic, it could well be the case that across the set of mosaics in the intersection of the constraints there is some variation  in the initial conditions of the universe and also in the numerical value of the parameters, so drawing a mosaic from the set entails arbitrarily selecting both initial conditions and parameter values. In this context both initial conditions and parameters are simply different facets of the degrees of freedom left by the laws of nature, and thus they have precisely the same modal status. Of course, it \emph{could} also be the case that the parameters are in fact fixed by the laws of nature while the initial conditions are not, but likewise it \emph{could} be the case that the initial conditions are in fact fixed by the laws of nature while the parameters are not: until we have some specific evidence for either parameter rigidity or state rigidity, parameters and initial conditions should be afforded the same modal status.

 \section{Objective Chance \label{objective}} 
 
Given the close links between determinism and objective chance, our proposed alternative approach to determinism will also have consequences for the concept of objective chance. Famously, objective chance is a murky notion - though a number of analyses of the topic exist, including frequentist and propensity approaches, none has so far garnered universal approval and all seem to have serious problems to overcome (\cite{SEPprobability}). It is tempting to respond to this situation by simply insisting that there aren't  any objective chances at the level of fundamental laws, but that strategy is hampered  the fact that the current scientific evidence seems to be pointing away from Laplacean determinism. It may appear that we are faced with a dichotomic dilemma -  either the world satisfies Laplacean determinism or there exist objective chances - and thus, since it doesn’t appear to be the case that quantum mechanics satisfies Laplacean determinism, it may seem that we have no choice but to accept the existence of objective chances.  

But our discussion of holistic determinism demonstrates that this division of the possibilities is too rigid, for it turns out that  we can postulate worlds which do not satisfy Laplacean determinism but which nonetheless satisfy holistic determinism, and therefore the probabilistic features of quantum mechanics do not force us to accept the existence of objective chance from the holistic, external point of view. There are a variety of ways in which such apparently probabilistic events might emerge within a   world satisfying holistic determinism - they will appear wherever the world contains some events  which depend on facts about reality that are not accessible to observers in the local region of those events - but for concreteness, in order to demonstrate how chances can emerge from a world satisfying holistic determinism we will henceforth focus on one particularly simple possibility: a law of the form `events of type A have outcome A with probability 0.8' may be translated to a law imposing a constraint of the form `the set of mosaics in which eighty percent of events of type E have the outcome A.'  Evidently for the large majority of mosaics in this set, if we are in that mosaic and we observe a sufficiently large number of events of type E we will find the relative frequency of outcome A approaches 0.8, so this frequency-based law can be expected to give rise to the characteristic behaviour associated with the original probabilistic law.

The existence of these sorts of global constraints is entirely compatible with strong, weak or delocalised holistic determinism, so one could equally well say either that this approach reduces objective chance to frequency constraints, or that it simply eliminates objective chance altogether. We could even  think of it as a way of reducing objective chances to subjective probabilities: the relevant events are sampled from  a set of events with prespecified relative frequencies, so the objective chances attached to these events simply describe a process of sampling without replacement akin to the common textbook example of selecting a ball from a jar containing a specified mixture of black and white balls. And of course the probabilities involved in sampling without replacement can be described in terms of Bayesian credence functions with partially unknown initial conditions, since we know the initial proportions of black and white balls but not the order in which they will be drawn.  Thus deriving chances from frequency constraints provides  a satisfying explanation of the close conceptual and mathematical relationship between objective chance and subjective probability: objective chances behave like subjective probabilities because they really are just subjective probabilities over a large and temporally extended domain.

Frequency constraints are similar to a view that has been advanced by \cite{Roberts2009-ROBLAF} under the name of \emph{nomic frequentism};  much of Roberts' discussion is also relevant here, so we won't dwell on the points of agreement, but in appendix \ref{diff} we discuss a few points of difference. Frequency constraints are also clearly related to  finite frequentism, which is which is the view that objective probabilities are by definition equal to the actual relative frequencies of the relevant sort of event across all of spacetime(\cite{SEPprobability}). But the frequency constraint approach avoids several of the main problems encountered by finite frequentism. For example, it has been objected that finite frequentism entails that probabilities can't be defined for processes which as a matter of fact occur only once or not at all (\cite{Hajek}). But this is not a problem for the frequency constraint view provided that we take a robust attitude to modality which maintains that  constraints are ontologically prior to the Humean mosaic - in that case a constraint exists and is well-defined regardless of how many occurrences of the event type in question actually occur. Another common objection to finite frequentism is that it has the result that probability depends on a choice of reference class, and in the case of macroscopic everyday events this can be very nontrivial, because different descriptions of the event may suggest different natural reference classes which can lead to very different probabilities (\cite{Hajek}). However, this is not a problem for the frequency constraint view provided that we take constraints to be objective and mind-independent, because the reference class is defined within the constraint and therefore there is always a definite fact about which reference class is relevant. Of course, there is still an epistemological question about how we as observers can decide which reference class actually appears in the real underlying constraint, and it's always possible that we will get it wrong and will thus make the wrong predictions, but there is no ambiguity in the definition of the objective chance itself. \cite{Roberts2009-ROBLAF} gives more examples of the advantages of frequency constraints over finite frequentism.

In order to see if frequency constraints can give rise to chances which behave in the way we would expect, we will consider the following desiderata for an account of objective chance:

   \begin{enumerate} 
	
	\item \emph{The Principal Principle:} Objective chances should satisfy the Principal Principle.
	
	\item \emph{Probabilistic confirmation:} We should be able to confirm facts about objective chances by means of observing relative frequencies.
	\item  \emph{Exchangeability:} Given a sequence of events which are not causally related to one another, the chance for a given sequence should depend only on the relative frequencies in that sequence and not on the order in which the events occur. 
	\item \emph{Counterfactual independence:} Given a sequence of  events which are not causally related to one another, the chance for the outcome of an individual event should not depend on the frequency of occurrence of other outcomes in the sequence. 
	
\end{enumerate} 

\subsection{The Principal Principle}

Do chances derived from frequency constraints  satisfy the  Principal Principle? Well, in accordance with the method of direct inference (\cite{McGrew}), if you know that you are in a Humean mosaic which belongs to the set of mosaics in which exactly eighty percent of As are Bs, and  you have  no other information, you should indeed set your credences for the next observed $A$ to be $B$ to eighty percent. Now in most real cases we will already have observed some As and therefore it is not quite the case that we are sampling at random from the entire class, so the correct probability to assign might actually be infinitesimally different from eighty percent. But provided that the class is large enough, this difference will be so small that it will make no practical difference to the way in which we use this probability in our reasoning processes and thus  for all practical purposes the Principal Principle does indeed pick out the frequencies that appear in the constraints.

\subsection{Probabilistic Confirmation} 

Next let us check for probabilistic confirmation: if we observe a set of events of a certain type and find that eighty percent of them have the outcome A, do we have grounds to confirm the hypothesis that we are in a Humean mosaic in which eighty percent of instances of this event type have outcome A? Of course, in general we will be able to observe only a very tiny proportion of the total number of such events occurring across the whole Humean mosaic, so we can't confirm this hypothesis by direct observation; rather we must assume that the As that we have observed are a \emph{representative sample} of the full set of As. Thus the 
legitimacy of this inference depends crucially on the assumption that the As form a homogenous class which can be expected to exhibit stable relative frequencies across time -  it is not the case that just any relative frequencies can be extrapolated in this way, as that would allow us to confirm virtually any hypothesis we like about future relative frequencies by simply labelling events in the right sort of way. Thus from observations of a relatively small subclass of instances of a given event type, we do not have grounds to confirm the hypothesis that  eighty percent of instances of this event type have outcome A, but we \emph{do} have grounds to confirm the hypothesis that the laws of nature induce a constraint to the effect that eighty percent of As must be Bs. That is, the inference must include the hypothesis that we have identified a special class of events which appears directly in the laws of nature and that the relative frequencies we have seen are the consequence of a constraint which is induced by the laws of nature. Of course we could always be wrong about this hypothesis, but nonetheless, observing behaviour which is consistent with the hypothesis provides some degree of confirmation for it. 
 
 \subsection{Exchangeability and Counterfactual Independence }
 
Let us move to the final two desiderata, which are somewhat more mathematical in character. It is clear that  chances obtained from frequency constraints will satisfy exchangeability, since these constraints govern only relative frequencies and not order of appearance. But what about counterfactual independence? Well, we have seen that  in the frequency constraint picture, observing random events must be understood as a form of sampling without replacement, and so just as we would update our probabilities for the next event after each random draw when performing sampling without replacement, it seems we ought to do the same in the frequency constraint case: if I know that exactly 50 percent of events of a given type must have the outcome A, then if my first one hundred observations all have the outcome B I will conclude that the remaining set of events must contain one hundred more As than Bs, so I will perform the standard sort of inference we see in sampling without replacement, which involves updating my credences to reflect the fact that the next outcome is more likely to be A than B. Note that this remains true for any finite total number of events, so the conclusion does not change if I don't know the total number of events. It also doesn't matter if we have a constraint which prescribes the probability only approximately; there will always be some threshold number $n$ of Bs such that if I have observed $n$ more Bs than As, then the set of unobserved events must contain more As than Bs in order to get  final frequencies which are correct to within the limits allowed by the relevant   constraint.

This seems problematic, as  the absence of counterfactual independence jars with many of our intuitions about probability - for example, it has the consequence that under certain circumstances the gambler's fallacy is not really a fallacy at all. Indeed Hajek argues against standard frequentism on similar grounds, pointing out that defining probabilities as relative frequencies prevents us from saying that the probability for an individual event is counterfactually independent of what happens in other similar events distant in space and time (\cite{Hajek}). So we must now ask ourselves whether we can   accept an approach which fails to satisfy counterfactual independence  as an acceptable analysis of objective chance. One possible response for the proponent of frequency constraints is to point out we will not get failures of counterfactual independence if the total number of instances is infinite, since sampling without replacement from an infinite set is mathematically equivalent to sampling \emph{with} replacement. And after all, we can never know for sure that the number of instances of an event type is finite, since that would require us to have illegitimate knowledge of the future, so it might be argued that we should never make updates to our probabilities which are premised on the number of instances of an event type being finite, meaning that the usual sort of inferences we see in sampling-without-replacement should not be allowed with regard to sequences of chancy events. However, although it is true that we can never know for sure that the number of instances is finite, nonetheless one can imagine cases where our best theories give us good reason to believe that this is so (for example, if we had a theory which implied that spacetime is discrete and the universe has a finite extent and time has both a start and an end), so at least in principle there are circumstances where counterfactual independence really would be violated in this picture. 

Another possible response is to point out that we would only ever get violations of counterfactual independence with regard to sequences of chancy events if we had knowledge of  frequency constraints that would trump the observed relative frequencies, and one might think that this could never come about. For example, in real life if we performed one hundred measurements and got the result B every time, we would probably form the working hypothesis that there is a constraint requiring that all events of this type must have outcome B, rather than supposing that exactly fifty percent of all events of this type must have outcome B and that therefore subsequent events must be more likely to have outcome A than outcome B in order to make up the total.  However, probabilistic laws do not usually exist in isolation, and therefore one can imagine a case where we decide based on considerations of symmetry or coherence with the rest of our theory that the objective chance of B is 0.5 even though so far we have seen significantly more Bs than As. It would seem that under those circumstances, if we accepted the frequency constraint analysis and we had reason to believe that the total number of instances of the relevant event type across all of spacetime would be finite, we would have reason to assign a credence greater than 0.5 to the proposition that the next such event will have outcome A.

So in fact, probably  the appropriate response here is to simply say that violations of counterfactual independence \emph{should} sometimes be allowed with respect to sequences of chancy events, because there is little reason to expect our intuitions to be a good guide on this point. For a start, most of our intuitions are based on probabilistic events that involve \emph{subjective} probabilities rather than objective chances of the kind that appear in the fundamental laws of nature - due to decoherence, macroscopic events do not typically depend sensitively on quantum mechanical measurement outcomes, and therefore paradigmatic probabilistic events like `rolling a die' or `flipping a coin' can be understood entirely in terms of our ignorance of the specific initial conditions. Provided that the relevant initial conditions are independent (or at least independent enough for practical purposes) these sorts of events do indeed obey counterfactual independence, which is likely the source of our intuitions on this point.  Furthermore, even when we do observe events that may involve objective chances, such as the results of quantum measurements, we are presumably observing only a very small subset of the total number of events of this type, so we should not expect these observations to tell us very much about the class as a whole. If we know that exactly fifty percent of instances of a certain event type have the outcome $A$, and also that $10^{11}$ of these events occur across all of spacetime, then in principle the fact that we have observed one hundred more $B$ outcomes than $A$ outcomes does make it more likely that the next outcome will be an A, but in practice the change in probability is so small that it will  not lead to any observable effects.

Therefore, the fact that counterfactual independence conflicts with our intuitions is no reason to reject  chances obtained from frequency constraints, since our intuitions have been developed in entirely the wrong context of application. Indeed, counterfactual independence can be regarded as a way of expressing the expectation that objective chances should be intrinsic properties  of individual entities which are spatially and temporally local; but in many ways  chances  actually look like properties not of individual entities but rather of large collections of entities. For example, we can't ascertain chances during a single observation, but rather we must observe a large number of events of the same type and then make an inference about what the chances are. A number of popular approaches to the analysis of objective chance - particularly the propensity interpretation, which holds that chances are intrinsic properties of individual entities akin to colours and masses (\cite{popper2005logic}) -  struggle to make sense of this feature (\cite{Eagle}). But frequency constraints do justice to the apparently collective nature of objective chance by turning chances into a form of global coordination across space and time, where events are constrained to coordinate non-locally amongst themselves so as to ensure that outcomes occur with (roughly) the right frequency.  \cite{Hajek} worries that `\emph{it's almost as if the frequentist believes in something like backward causation from future results to current chances,}' - and indeed, that is exactly what the proponent of frequency constraints does believe!

\section{Related Topics \label{rt}}

In this section, we discuss the relationship between holistic determinism and some other relevant research programmes. 

\subsection{Superdeterminism} 

Since the terms `holistic determinism' and `superdeterminism' are superficially similar, it's important to reinforce that these concepts are not identical. Broadly speaking, `superdeterminism' describes approaches to quantum mechanics which deny the existence of non-locality in quantum mechanics by rejecting the assumption of statistical independence which goes into Bell's theorem, i.e. the idea that our choice of measurement is independent of the state of the system which we are measuring (\cite{10.3389/fphy.2020.00139,hossenfelder2020superdeterminism,palmer2016invariant}). There are two typical ways to achieve this. First, given that our choice of measurement is always determined or at least influenced by facts about the past (e.g.  the physical state of our brain in the time leading up to our choice) one could imagine the initial state of the universe being arranged such that our measurement choices are always correlated with the states of the systems we are measuring. Second, one could imagine there is some sort of retrocausality or `future outcome dependence' where the state of the system we are measuring is influenced by our future decisions about what measurement to perform. 

It is immediately clear that neither of these approaches actually requires determinism: a \emph{probabilistic} dependency between choice of measurement and state is already enough to vitiate statistical independence and thus restore locality. Thus it is actually somewhat misleading to use the term `superdeterminism' to refer to the violation of statistical independence. The term `superdeterminism' contains the word `determinism' because Bell originally proved a theorem ruling out deterministic local hidden variable theories (\cite{Bell}), and so at that time superdeterminism was suggested as an approach to restore both locality and determinism; but a later version of Bell's theorem ruled out \emph{all} local hidden variable theories, either probabilistic or deterministic (\cite{articleEsfeld}), so in fact determinism is not really the key issue here, and modern proponents of superdeterminism are usually motivated more by the desire to preserve locality than any particular interest in preserving determinism. 

That said, for the moment let us focus on  superdeterministic approaches which \emph{are} also deterministic in some sense. What sort of determinism would that be?  Well, if we take the first route where correlations between measurement choices and states are written into the initial state of the universe, then we have just standard Laplacean determinism (and thus also weak holistic determinism). If we take the second route involving retrocausality, the resulting theory would probably not be compatible with Laplacean determinism, because events which are determined by something in the future usually can't also be fully determined by the past; but such a theory certainly could exhibit either weak or strong holistic determinism. However, although superdeterminism can be a form of holistic determinism, the concept we have defined here is significantly more general than superdeterminism. In particular, a major motivation for our approach to was to understand what determinism could look like in a world which includes spatial and/or temporal non-locality, whereas superdeterminism is typically employed as a way of \emph{denying} the existence of either spatial or temporal non-locality, and therefore holistic determinism also accommodates realist approaches which take a very different approach from superdeterminism.

\subsection{Deterministic Probabilities}

There is a long-standing philosophical debate around the question of whether it is possible to have chances in a deterministic world (\cite{McCoyManuscript-MCCAOG-3,Glynn2010-GLYDC,GallowForthcoming-GALASG}), but this debate usually takes place against the backdrop of Laplacean determinism. Moving to a picture based on holistic determinism therefore opens up new ways in which we can have `chancy' events in a deterministic world, since we saw in section \ref{objective} that an event can be described by a non-trivial probability distribution when we condition on all of the information available to local observers, even if the event is not objectively chancy when we consider the way in which it is embedded in the complete objective modal structure of reality. This reinforces the point that `chanciness' is in many cases relative to a perspective: obviously if the outcome of a process is determined by something in the future it is for all practical purposes objectively chancy from the point of view of observers like ourselves who have directly epistemic access only to the past and present, but this does not necessarily mean that the process must be `chancy' from the point of view of the underlying modal structure.

\section{Conclusion} 

We have argued that the metaphysical picture of determinism captured in Laplace's original vision is no longer fit for purpose, since it can't be usefully applied to the diverse range of laws outside the time evolution paradigm that appear in modern physics. However, we contend that even for a world governed by laws which are not time evolution laws, it is still meaningful to ask  whether or not that world is deterministic.  Thus in this article we have provided several generalized definitions of determinism which can be applied to a wide variety of nonstandard laws, thus rehabilitating determinism for the post-time-evolution era.

 Using these definitions, we have shown that a world governed by laws outside the time evolution paradigm may fail to satisfy the standard conception of Laplacean determinism but may nonetheless exhibit a form of determinism on a global scale, where we say we have  some form of `determinism' provided that we are never required to assign probability distributions over entire Humean mosaics or worlds. This has the consequence of dissolving the traditional dichotomic dilemma where we supposedly have to choose between Laplacean determinism and intrinsic randomness - a world governed by global aws may exhibit what appear to be objective chances from the point of view of local observers whilst requiring no irreducible objective chances from the external, holistic perspective. We have discussed one possible way of implementing this vision by means of constraints prescribing relative frequencies across spacetime, although frequency constraints are only one possible way in which chancy events could arise within a world satisfying holistic determinism.
 
 Several interesting conceptual points have come up along this journey. In particular, the possibility of weak holistic determinism requires something of a conceptual shift, requiring us to distinguish between events which are chancy and events which are undetermined but nonetheless not chancy. This distinction is not entirely new -  for example, it is common to hold that the initial conditions of the world are not determined by anything but also not governed by any objective probabilities - but it is not always well articulated, and often fails to be taken seriously as an option for any phenomena other than the initial conditions of the universe.   Weak holistic determinism thus offers a welcome halfway house for those who find it implausible that everything in the universe is fully determined by the laws of nature, but who are also suspicious of the ill-defined concept of objective chance. We have briefly surveyed some of the consequences of this new perspective, nothing that it offers interesting insight into for long-standing  debates in which determinism plays a role within both physics and philosophy.

 \appendix

 \section{Nomic frequentism \label{diff}}
 
The `nomic frequentism' proposal of \cite{Roberts2009-ROBLAF}, which suggests that laws about probabilities should be understood in terms of laws of the form `R percent of the Fs are Gs'  has much in common with the frequency constraint approach that we have discussed in this article. However, there are a few key differences.  First, Roberts presents nomic frequentism as a general analysis of all probabilities that appear in laws, including for example the laws of evolutionary biology, whereas we have suggested it only for the specific case of objective chances appearing in the fundamental laws of nature, and in particular quantum mechanics - we would expect that the laws of evolutionary biology could be satisfactorily understood in terms of subjective probabilities. Moreoever, we do not even claim that \emph{all} of the objective chances appearing in the fundamental laws of nature must be attributed to frequency constraints; we merely observe that this is one possible way in which apparently probabilistic events could arise in a deterministic universe.  So our account is in that sense considerably less general and ambitious than Roberts'. 

Second, Roberts suggests that `eighty percent of As are Bs' should be regarded as conceptually equivalent to a law like `All As are Bs,' (or rather, the latter should be regarded a special case of the former). But there is one important conceptual difference: the constraint `All As are Bs' does not seem to require any `communication' between the As, as it is enough that each $A$ has the intrinsic property of being a $B$, whereas `exactly eighty percent of As are Bs' or even `approximately eighty percent of As are Bs' does seem to require some sort of coordination, as in order to ensure that the relative frequency is exactly or approximately correct, it seems that each A must `know' something about what the other As are doing. Thus frequency constraints seem to require some form of non-locality (as Roberts himself later observes), which means we are in quite a different conceptual space from standard laws like `all As are Bs.'

Third, Roberts' solution to the apparent absence of counterfactual independence within nomic frequentism is  to say that `the way a nomic frequentist will represent independence of distinct fair coin-tosses is by denying the existence of any law that implies that the conditional frequency of heads on one toss given the results of another toss is different from (50 percent).'  But as we have seen, it does not seem that this can be entirely correct: constraint frequentism does allow the violation of counterfactual independence in at least certain special cases. Roberts justifies his solution on the basis that knowing facts which violate counterfactual independence `would require (us) to have advance intelligence from the future,' but this argument seems to depend implicitly on the assumption that we are in a Humean context where the only information which is relevant to  inferences about a future event is an actual observation of the event itself. But if the laws which induce the   frequency constraints are understood as laws which are ontologically prior to the Humean mosaic, and if we are able to make correct inferences about the laws based on observations of a limited subset of the mosaic, then we would in principle be able to know about violations of counterfactual independence without having any illegitimate information about events which have not yet occurred. This sort of `knowledge of the future' is not really any different from the knowledge of the future that we get from more familiar sorts of scientific laws:  the laws constrain the Humean mosaic, including the future, and thus by figuring out the laws we can make inferences about the future.

  \bibliographystyle{plainnat}
 \bibliography{newlibrary12}{}

\end{document}